%% file: arXiv_submitting.tex
\documentclass[useAMS,usenatbib]{mn2e}

\usepackage{amsmath}
\usepackage{amssymb}
\usepackage{graphicx}
\usepackage{subfigure}
\usepackage{color}

\title[Exploration for PBHs using PTAs]{Enhanced exploration for  primordial black holes using pulsar timing arrays}
\author[K. Kashiyama,  N. Seto]{Kazumi. Kashiyama$^{1,2}$\thanks{E-mail:kzk15@psu.edu},  
Naoki. Seto$^{3}$\\
$^{1}$Department of Physics and Center for Particle Astrophysics, Pennsylvania State University, University Park, PA, 16802\\
$^{2}$Department of Astronomy $\&$ Astrophysics,, Pennsylvania State University, University Park, PA, 16802\\  
$^{3}$Department of Physics, Kyoto University, Kyoto 606-8502, Japan}
\begin{document}

\date{Accepted 2012 August 14. Received 2012 June 14}

\pagerange{\pageref{firstpage}--\pageref{lastpage}} \pubyear{2012}

\maketitle

\label{firstpage}

\begin{abstract}
We investigate the capability of pulsar timing arrays (PTAs) as a probe of primordial black holes (PBHs), 
which might constitute the Galactic dark matter. 
A PBH passing nearby the Earth or a pulsar gives an impulse acceleration and induces residuals on otherwise orderly pulsar timing data. 
We show that  the timing residuals induced at pulsars are optimal for searching 
heavier PBHs than those at the Earth, 
and the two probes are highly complemental. 
Future facilities like SKA could detect PBHs with masses around $\sim 10^{22\mbox{-}28} \mathrm{g}$ 
even if only a small fraction ($\lesssim 1\% $) of the Galactic dark matter consists of these PBHs. 
\end{abstract}


\section{introduction}
A significant fraction of matter in our Galaxy is considered to be occupied 
by dark matter, but  its nature is poorly understood at present 
\citep{2005PhR...405..279B}. Primordial black 
holes (PBHs)  are an interesting astrophysical candidate of dark matter, and 
various observational constraints  have been posed on their allowed mass range 
\citep{2010PhRvD..81j4019C,2010RAA....10..495K}.
However, the 
current constraints  in the mass range $10^{20}{\rm g}<M_\mathrm{PBH}<10^{27} {\rm 
g}$  remain relatively  weak.

To examine PBHs in this range, \cite{2007ApJ...659L..33S} proposed an 
  observational technique to directly probe
  gravitational interaction between the solar system and a nearby PBH. 
Around their close approach,  the Earth receives an impulse of acceleration whose profile
depends on the mass, distance, and velocity of the PBH.
Given the local mass density of dark matter, the expected magnitude of the 
acceleration is very small, but   
high-precision  measurements of  pulsar timing  could allow us to detect the weak 
impulse signal (see also \cite{2004PhRvD..70f3512S,2009PhRvL.102p1101S,Griest:2011}).

Roughly speaking, in the present context,  the pulsar timing analysis can be 
essentially regarded as  
measurement of the arrival times of radio pulses that were emitted by a pulsar and 
received on the Earth.  The local accelerations of both of the pulsar and the 
Earth are encoded in the modulation of the time-of-arrival (TOA) data, as 
a simple linear combination of two separate terms for the two masses.  We call 
them by the pulsar term and the Earth term respectively.

When observing multiple pulsars for  PBH search, the Earth terms are commonly 
excited by  
  the  acceleration of the Earth, and have coherent structure among TOA 
  data of different pulsars. Therefore, we 
can statistically amplify the weak impulse signature on the Earth by using a 
pulsar timing array (PTA) and  effectively 
reducing the timing noises.  The underlying statistical approach here is similar to  
that for detecting gravitational waves (GWs) whose effect on the TOA data can be 
also expressed with  two terms induced  separately  at the pulsar and the Earth 
 \citep{1978SvA....22...36S,1979ApJ...234.1100D,1983ApJ...265L..39H}.   
For the  GW signals, we call them by the pulsar GW term and the Earth GW term, 
in order to distinguish them from the two acceleration terms.  The traditional 
target of PTAs is the Earth GW terms.

Recently, it has been actively discussed that we  might utilize the pulsar GW terms 
for analyzing GW sources such as  merging super-massive black hole binaries  
\citep{2004ApJ...606..799J,2010arXiv1008.1782C,2011MNRAS.414.3251L,Ellis:2012}. 
 To this end, we need  
sufficiently stable millisecond pulsars (MSPs), and such preferred systems might be 
discovered   with future 
observational facilities.  With the pulsar GW
terms and known distances to the individual pulsars, our information content on GWs 
 at the nano-Hertz band can be greatly increased, compared with that extracted  
  from  the Earth GW terms alone.

 Meanwhile,  if the timing noises of individual pulsars are small, 
 we can also probe PBHs around the pulsars through the pulsar (acceleration) 
 terms, in addition to PBHs close to the solar system. Then, the pulsar terms can largely 
 widen our survey volume, and 
 our sensitivity for the direct PBH search would become better than the previous 
 estimation only with the Earth terms \citep{2007ApJ...659L..33S}. 
 In  this paper, 
we study this issue with special attention to the differences between the roles 
of two terms for the PBH search with PTAs, namely, the independent and numerous pulsar 
terms and the coherent  
 Earth terms. 
We find that these terms have both advantages and disadvantages for PBH search, and work in a complementary way.

\section{Probing primordial black holes using pulsar timing array}
\begin{figure}
\includegraphics[width=80mm]{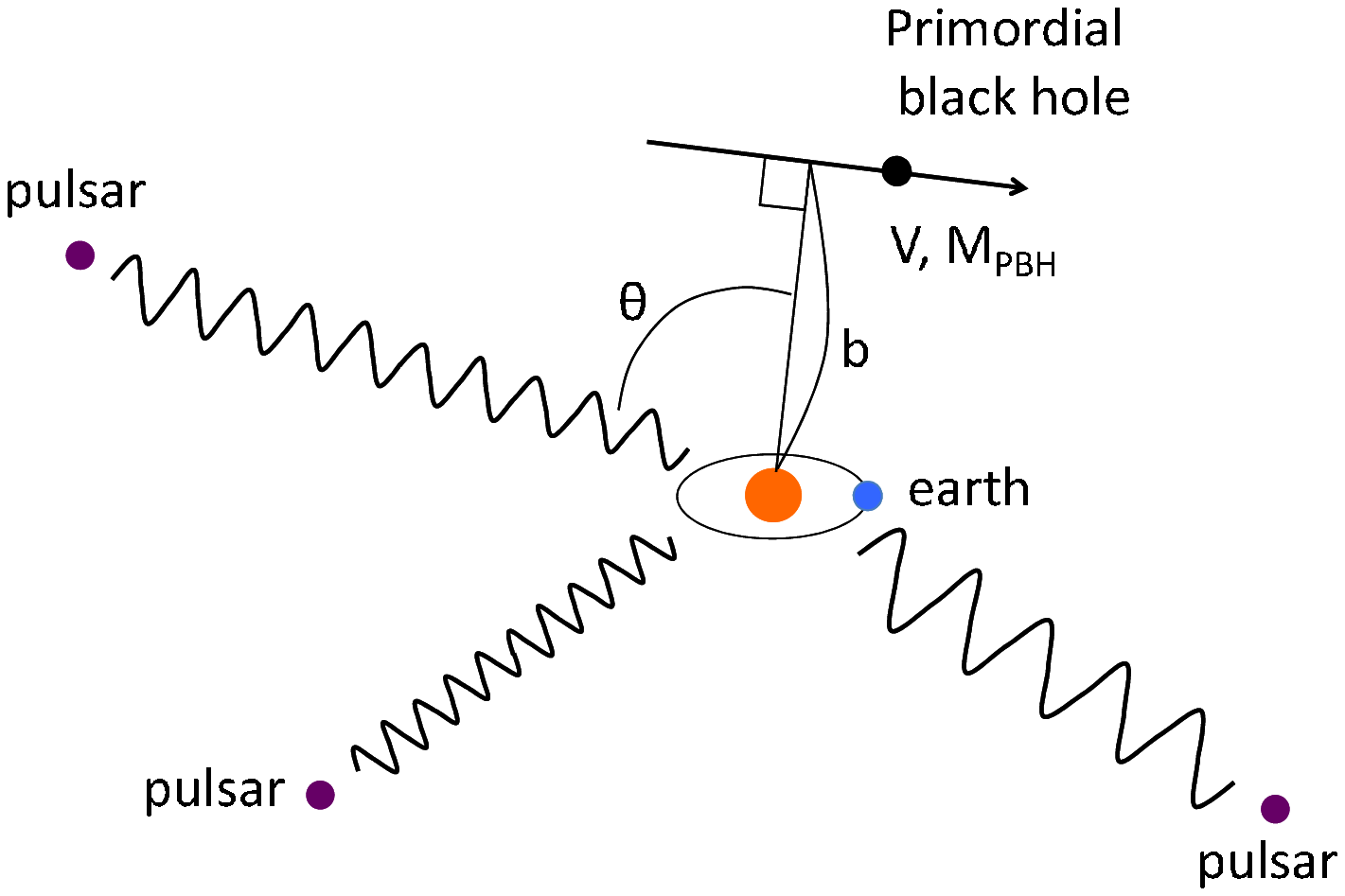}
\includegraphics[width=80mm]{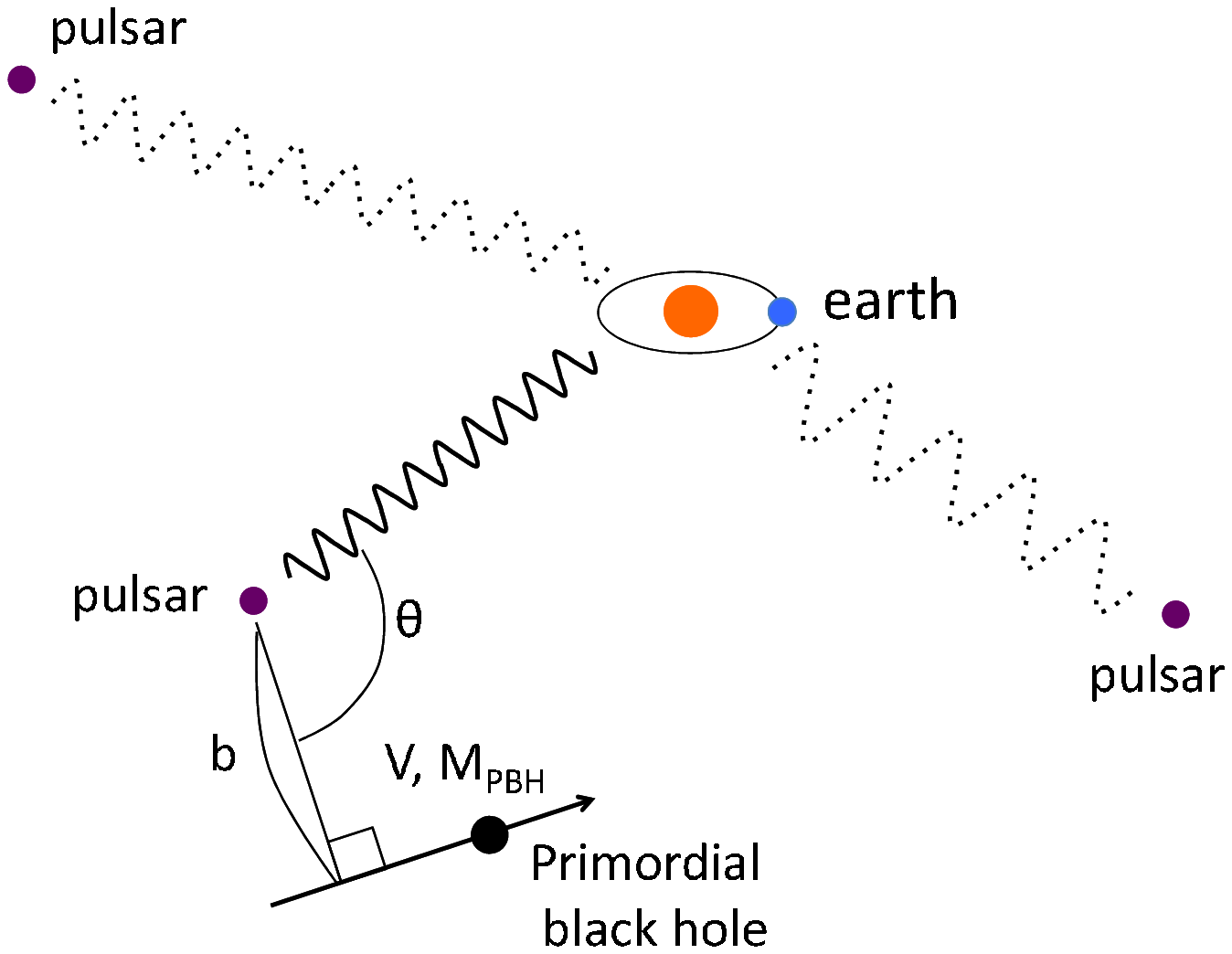}
\caption{Schematic pictures of primordial black hole (PBH) searches using a 
pulsar timing array (PTA). The impulse signal of a PBH is characterized by its 
mass  $M_\mathrm{PBH}$, the relative velocity $V$  to the target, the impact 
parameter $b$, and the projection angle $\theta$ between the pulsar-Earth line and the 
closest approach. {\it Top}  
panel shows the case with a PBH passing nearby the Earth. 
The acceleration by the PBH is imprinted in all the timing data available. By 
taking the correlation of the timing data, the signal can be  effectively amplified by a 
factor of 
$\sqrt{N_\mathrm{PSR}}$. {\it Bottom} panel shows the case with a PBH 
passing near a pulsar. The PBH modulates the timing data of the specific 
pulsar alone. The rate of such encounter is proportional to the number of pulsars $N_\mathrm{PSR}$.}
\end{figure}

In this paper, we assume that our PTAs are composed by totally  $N_\mathrm{PSR}$ pulsars with 
roughly the same level of timing noises. We further assume that the noises are 
white and have no correlation between different pulsars (see 
\cite{2010ApJ...725.1607S} for impacts of red noises).   
Note that  
the total numbers of the Earth  terms and the pulsar 
terms are both  $N_\mathrm{PSR}$.

In Fig.1,  we provide  schematic pictures of PBH search with a PTA.  
The top panel shows the case in which a PBH passes nearby the solar system  and 
excites the Earth terms coherently for  the TOA data of all the pulsars.
In the bottom panel  a PBH flybys a pulsar and modulates only the 
specific pulsar term.

 As mentioned earlier, we can amplify the sensitivity of the Earth 
terms by  a factor of $\sim 1/\sqrt{N_\mathrm{PSR}}$ using their coherent structure. 
In contrast, the flybys of  PBHs around the whole pulsars occur $\sim 
N_\mathrm{PSR}$ times more frequently than those around the Earth alone.

We hereafter assume that, when a PTA is available, the coherent Earth term can be 
removed from TOA data of each pulsar, and its pulsar term can be analyzed separately.

\subsection{Signal to noise ratio}
Here we estimate the signal-to-noise ratio (S/N) of a flyby event in the TOA 
data of a PTA.  The target mass can be the Earth or one of the   pulsars.
In the Fourier space, the main contribution of the impulse acceleration is the  mode with the frequency $f = 1/T$,  
where $T$ is the time scale for the PBH passing around the target. In the same 
manner,   the amplitude  $s_f$ of the mode can be estimated as 
$s_f \approx (a T^2/c) \times | \cos\theta |$.
Here $a$ is the peak magnitude of acceleration and $\theta$ is the angle 
between the pulsar-Earth line and the closest approach (see Fig.1).
With the  impact parameter $b$ and the relative velocity $V$,  the two quantities
$T$ and $a$ are given as
\begin{equation}\label{duration}
T \approx \frac{b}{V} \sim 10 \mathrm{yr} \left( \frac{b}{740 \mathrm{AU}}\right)  \left( \frac{V}{350 \mathrm{km/s}}\right)^{-1},
\end{equation}
and $a \approx GM_\mathrm{PBH}/b^2$, where $M_\mathrm{PBH}$ is the mass of the PBH.
Then the Fourier amplitude of the timing residual is  expressed as $s_f \approx (G M_\mathrm{PBH}/cV^2) \times |\cos\theta| $, or
\begin{equation}\label{timing_residual}
s_f \sim 10 \mathrm{ns} \left( \frac{M_\mathrm{PBH}}{10^{25} \mathrm{g}}\right) \left( \frac{V}{350 \mathrm{km/s}}\right)^{-2} \left( \frac{| \cos \theta |}{0.58} \right).
\end{equation}

Next we evaluate the noise associated with the  
 timing analysis. 
Using  the sampling rate $\nu$ of TOAs 
 and the rms noise $\sigma$ of each TOA, the Fourier mode of the timing noise at the frequency $f=1/T$
 is given as  $\sim \sigma/\sqrt{T \nu}$.  This result is  for TOA data of a single pulsar.
When we deal with a PTA, it is important to distinguish whether 
 we examine the Earth terms or a pulsar term.  For the former, we have the 
 statistical reduction factor $1/\sqrt{N_\mathrm{PSR}}$ from the coherence of the signals.
Therefore, the effective noise level for the impulse search  can be expressed as
\begin{equation}\label{timing_noise}
n_f \sim 6.2 \mathrm{ns} \left( \frac{\sigma}{100 \mathrm{ns}} \right) \left( \frac{T}{10\mathrm{yr}} \right)^{-1/2} \left( \frac{\nu}{0.50 \mathrm{wk}^{-1}} \right)^{-1/2}N_\mathrm{PSR}{}^{-E/2}
\end{equation}
with $E = 0 \ \mathrm{or} \ 1$  for a pulsar term and the Earth terms, 
respectively. From Eqs.(\ref{timing_residual}) and (\ref{timing_noise}) the 
signal-to-noise ratio of the flyby detection is now given as S/N $\equiv s_f/n_f$.

The signal of PBH is observationally characterized by the duration  $T$ and the amplitude $s_f$.  
On the other hand, there are four physical parameters, $b$, $M_\mathrm{PBH}$, 
$V$, and $\theta$.\footnote{For a PBH passing around the Earth, we can, in principle, estimate the 
direction of its closest approach, using the dipole pattern of the Earth terms 
in a PTA. } 
Thus, in general, we can obtain only two constraints between these four parameters,
even if the signal is detected with a high signal-to-noise ratio. 
However, when constraining the parameters of PBHs as a dominant component of dark matter, 
we can  set  fiducial values for $V$ and $\theta$ based on the following considerations.  
The rms velocity for halo dark matter relative to the solar system is dynamically estimated to be $\sim 350 \mathrm{km/s}$ \citep{Carr:1999}. 
Besides, the typical peculiar velocity of the observed MSPs are relatively small $\lesssim 100 \mathrm{km/s}$ \citep{2005MNRAS.360..974H}. 
Thus, it is reasonable to fix $V = 350 \mathrm{km/s}$ for discussing both the Earth and pulsar terms induced by PBHs. 
Furthermore, if the scatterings between PBHs and the targets occur isotropically, 
 the ensemble average of the projection angle $\theta$ becomes
$\sqrt{ \langle \cos^2\theta \rangle } = 1/\sqrt{3} \sim 0.58$. 
Hereafter we   fix $|\cos \theta|=0.58$ as  the fiducial value.  
Now the observational parameters ($T$ and $s_f$) and the physical parameters 
($b$ and $M_\mathrm{PBH}$) have one-to-one correspondence in our 
order-of-magnitude estimation. 
\footnote{We should note that some of the observed MSPs have comparable or larger peculiar velocities ({\it e.g.} B1957+20 \cite{2005MNRAS.360..974H}). 
Also the scattering between PBHs and the targets may occur in a non-isotropic way, in which $\sqrt{ \langle \cos^2\theta \rangle}$ takes a different value. 
We discuss these cases in the final section.}

\subsection{Event rate}

The density of the local dark matter is estimated to be $\rho_\mathrm{DM}=0.011 
M_\odot \mathrm{pc}^{-3}$ (\cite{2001MNRAS.326..164O}, see also 
\cite{2012arXiv1204.3924M} for a recent claim).
We put $\eta$ as the mass fraction of PBHs among the Galactic dark matter, and 
assume that the PBHs have an identical mass parameterized by $M_{\mathrm{PBH}}$.
\if
When we assume that a fraction $\eta$ of dark matters is consist of PBHs, the density of PBH can be described as 
$\rho_\mathrm{PBH} = \eta \rho_\mathrm{DM}$. 
\fi
Then we can estimate the event rate of the close encounter of a PBH around the target masses 
within a impact parameter $b$ as 
$R \approx  \pi b^2 V \times (\rho_\mathrm{PBH}/M_{\mathrm{PBH}}) \times N_\mathrm{PSR}{}^{1-E} $, or 
\begin{eqnarray}\label{event_rate}
R &\sim& 0.032\mathrm{yr}^{-1} \left(\frac{\eta}{1}\right) \left( \frac{\rho_\mathrm{DM}}{0.011 M_\odot \mathrm{pc}^{-3}}\right)  \left( \frac{M_\mathrm{PBH}}{10^{25} \mathrm{g} }\right)^{-1} \notag \\ &\times& \left( \frac{b}{740 \mathrm{AU}}\right)^2 \left( \frac{V}{350 \mathrm{km/s}}\right)  N_\mathrm{PSR}{}^{1-E}.
\end{eqnarray}
We take $E = 0 \ \mathrm{or} \ 1$ depending on whether a PBH flybys one of the  
pulsars ($E=0$) or the Earth ($E=1$).
The factor $N_\mathrm{PSR}{}^{1-E}$ reflects the fact that,  
only in the case of using the independent pulsar terms,  the event rate is proportional to the number of pulsars.

\subsection{Detectable parameter regions}

Now we discuss the observational prospects of  PBH search.  To this end, we 
examine the detectable PBHs in the two dimensional space $(M_{\mathrm{PBH}},b)$, 
  assuming two future PTAs  whose basic parameters are summarized in Table.1.
PTA$a$ has a relatively conservative set of  parameters, and could be realized 
in the near future  \citep{Ellis:2012}.  
The goal of PTA$b$  is  more challenging and might be realized with future 
facilities like SKA ({\it e.g.} \cite{Smits:2011}). The latter
will also enable us  
to estimate the parallax distances to the stable pulsars with typical 
errors of  $\lesssim 20\%$. Then we can obtain broad scientific results,  including a 
map of  the interstellar 
electron density \citep{Smits:2011}. 
\begin{table}
\caption{Parameter sets of the two pulsar timing arrays assumed for future 
prospects of PBH searches}
\label{tab_1}
\renewcommand\arraystretch{1.2}
\begin{tabular}{c|c|c}
 & PTA$a$ & PTA$b$ \\ 
\hline 
timing noise in each TOA ($\sigma$) & $100 \mathrm{ns}$ & $10 \mathrm{ns}$ \\ 
frequency of TOA sampling ($\nu$) & $0.5 \mathrm{wk}^{-1}$ & $1.0 \mathrm{wk}^{-1}$ \\ 
number of pulsars ($N_\mathrm{PSR}$) & $100 $ & $ 1000 $ \\ 
observation time ($T_\mathrm{obs}$) & $10 \mathrm{yr}$ & $20 \mathrm{yr}$ \\ 
\hline
\end{tabular}
\end{table}

For detecting  a PBH, we request that the following three conditions 
are simultaneously satisfied in the  $(M_{\mathrm{PBH}},b)$-plane; \newline
(i) S/N is larger than a threshold value ({\it e.g.} S/N $> 3$ for $99\%$ 
confidence level). \newline
 (ii) At least one event occurs during the observation time ($R \times 
 T_\mathrm{obs}  
 > 1$). \newline
(iii) The duration of signal is shorter than the observation time ($T < T_\mathrm{obs})$.

From Eqs.(\ref{timing_residual}) and (\ref{timing_noise}), the condition (i) 
corresponds to the region in the  $(M_{\mathrm{PBH}},b)$-plane  as 
\begin{eqnarray}\label{condition_i}
\left(\frac{M_\mathrm{PBH}}{10^{25} \mathrm{g}}\right) \left(\frac{b}{740 \mathrm{AU}}\right)^{1/2} &\gtrsim& 0.56 \left( \frac{\sigma}{100 \mathrm{ns}}\right)   \left( \frac{\nu}{0.5 \mathrm{wk}^{-1}}\right)^{-1/2} \notag \\ &\times&\left( \frac{V}{350 \mathrm{km}}\right)^{5/2} \left( \frac{| \cos \theta |}{0.58} \right)^{-1} \notag \\ &\times& \left( \frac{\mathrm{S/N}}{3}\right) N_\mathrm{PSR}{}^{-E/2} .
\end{eqnarray}
By reducing the rms noise $\sigma$, 
the overall 
sensitivity is improved and smaller PBHs are within reach. 
With the Earth terms,  the detectable   PBH mass becomes  
$1/\sqrt{N_\mathrm{PSR}}$ time smaller  than the limit with a pulsar term.

From Eq.(\ref{event_rate}), the condition (ii) can be expressed as 
\begin{eqnarray}\label{condition_ii}
 \left( \frac{M_\mathrm{PBH}}{10^{25} \mathrm{g}}\right) \left( \frac{b}{740 \mathrm{AU}}\right)^{-2} &\lesssim& 0.32  \left( \frac{\eta}{1}\right) \left( \frac{\rho_\mathrm{DM}}{0.011 M_\odot \mathrm{pc}^{-3}}\right) \notag \\ &\times&  \left( \frac{V}{350 \mathrm{km/s}}\right) \left( \frac{T_\mathrm{obs}}{10 \mathrm{yr}}\right) \notag \\ &\times&  N_\mathrm{PSR}{}^{1-E}.
\end{eqnarray}
When fixing the PBH fraction $\eta$, the event rate scales as  $R \propto 
M_\mathrm{PBH}{}^{-1}$,  and PBHs encounter the individual target 
masses less  
frequently for larger $M_\mathrm{PBH}$.   
On the other hand, the event rate is also proportional to the number of the 
target masses.   
Thus,  rare encounters with  massive PBHs can be probed with the  pulsar terms.

From Eq.(\ref{duration}), the condition (iii) for the signal duration can be expressed as
\begin{equation}\label{condition_iii}
b \lesssim b_\mathrm{max} = 740 \mathrm{AU} \left(\frac{T_\mathrm{obs}}{10 \mathrm{yr}}\right) \left( \frac{V}{350 \mathrm{km/s}} \right)^{-1}.
\end{equation}
When we fix the relative velocity, the signal duration $T$ has the one-to-one 
correspondence to the impact parameter $b(\equiv VT)$. 
The observation time limits the duration of the  detectable signal, and so does the impact parameter.    

\begin{figure}
  \includegraphics[width=80mm]{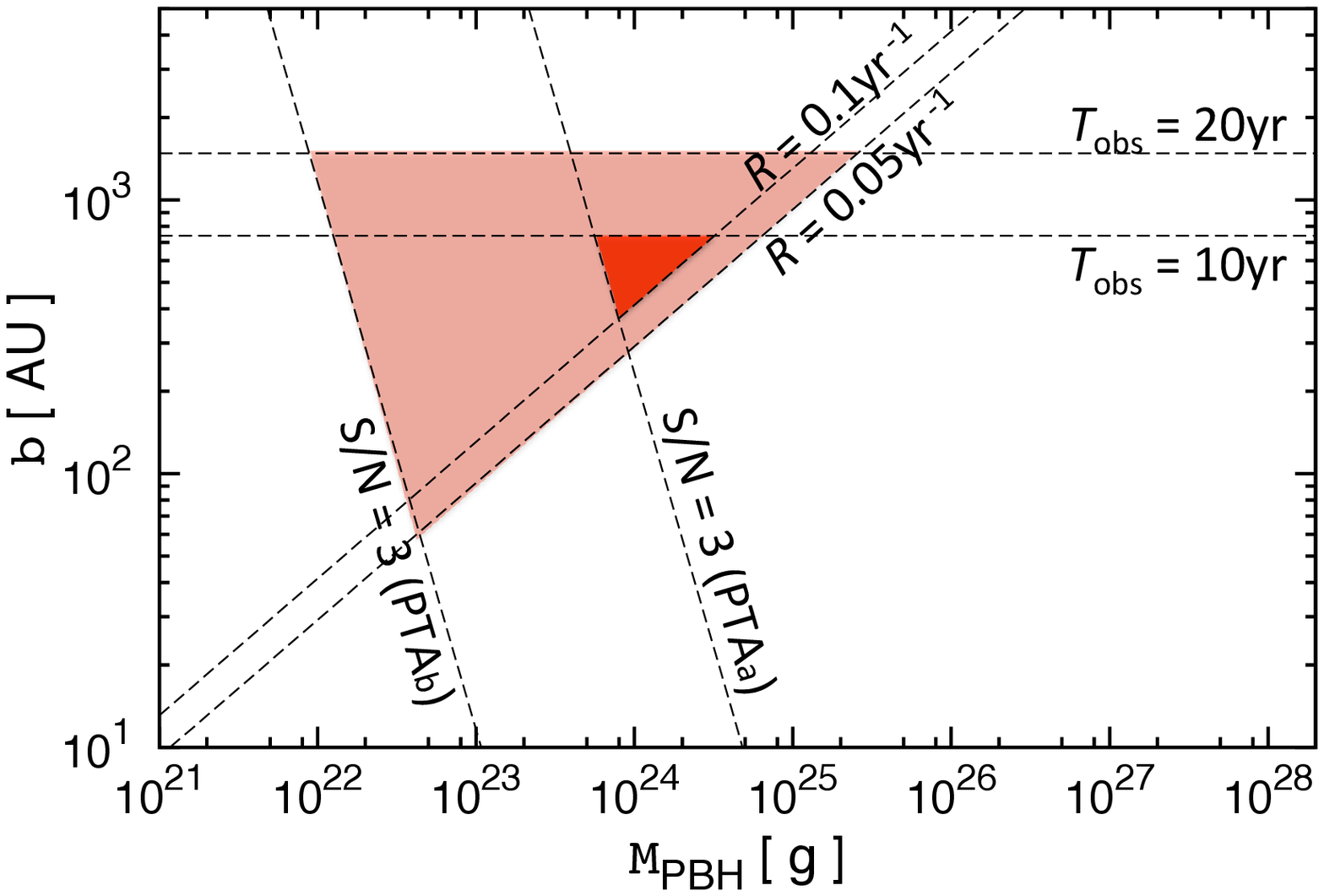} 
  \includegraphics[width=80mm]{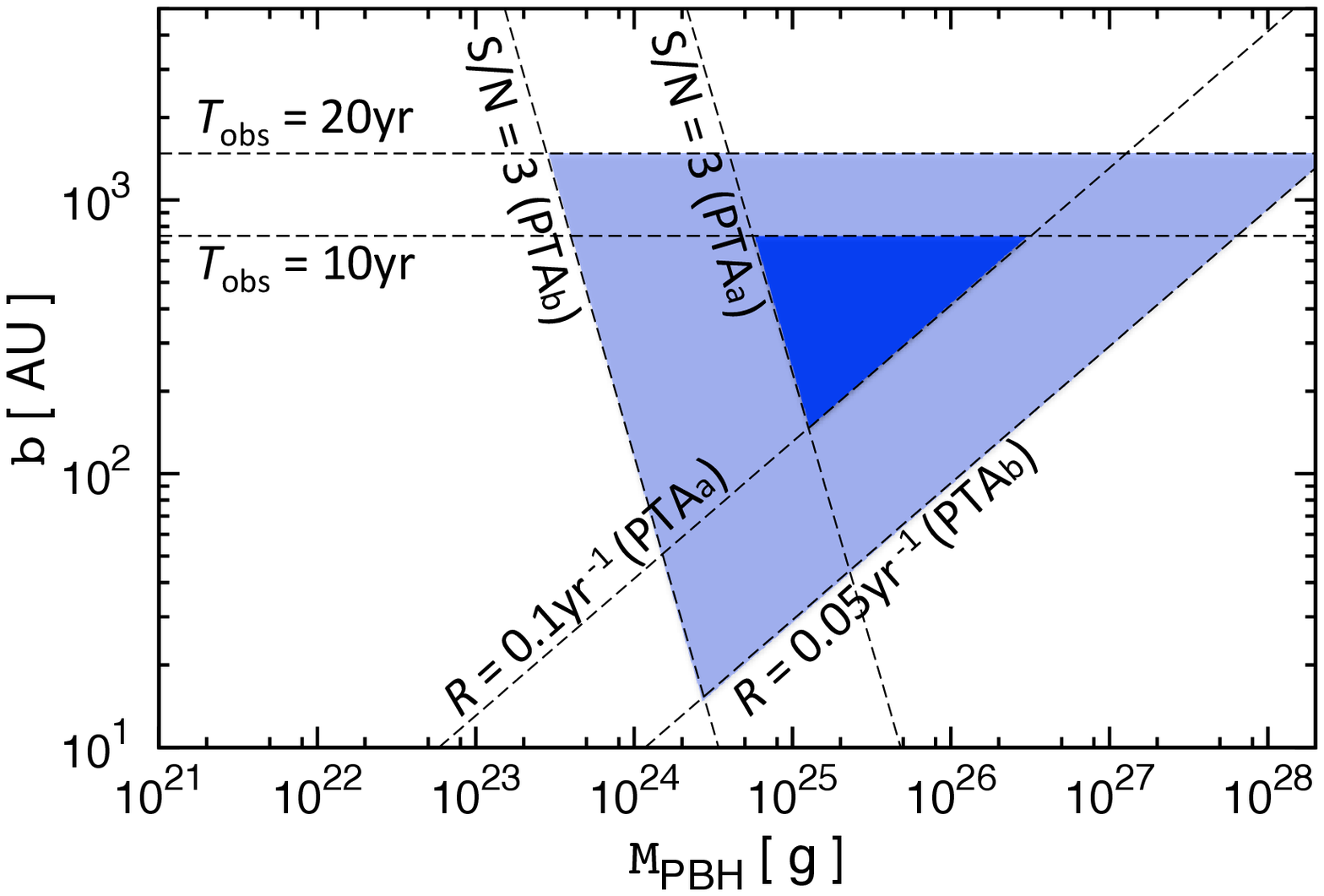}
  \caption{Detectable parameter regions of a primordial black hole (PBH) with 
mass $M_\mathrm{PBH}$ passing nearby the Earth (top panel) or a pulsar (bottom 
panel) with impact parameter $b$. The dark-colored regions correspond to the 
detectable events by PTA$a$, and the light-colored ones for PTA$b$ 
(see Table.1). Here we fix the relative velocity of  PBHs  at 
$V = 350 \mathrm{km/s}$, and the projection factor at $|\cos \theta | = 
0.58$. We assume that  the dark matter density is $\rho_\mathrm{DM} = 0.011 
M_{\odot} \mathrm{pc}^{-3}$ with the PBH fraction $\eta = 1$.  The lines with 
the slopes  $b \propto M_\mathrm{PBH}{}^{-2}$, $b \propto 
M_\mathrm{PBH}{}^{1/2}$, and $b = \mathrm{const}$   correspond to the conditions 
(i), (ii) and (ii) in the main text, respectively. }
\end{figure}\label{fig_2}

Fig.2 shows detectable parameter regions of a PBH in the
$(M_\mathrm{PBH},b)$-plane for $\eta=1$. The top panel is for a PBH passing near the Earth, and 
the bottom one is for those  near the pulsars. 
In each panel, the dark-colored regions represent the detectable events with 
PTA$a$, and the light-colored ones are for PTA$b$. 
 
In Fig.2, the lines with $b \propto M_\mathrm{PBH}{}^{-2}$, $b \propto M_\mathrm{PBH}{}^{1/2}$, and $b = \mathrm{const}$ 
correspond to the conditions (i), (ii) and (iii), respectively (see Eqs.(\ref{condition_i}), (\ref{condition_ii}) and (\ref{condition_iii})).
As one can see easily from Fig.2, the Earth terms and the pulsar terms are highly complemental. 
The Earth terms cover a lighter mass range $\sim 10^{22\mbox{-}24} \mathrm{g}$, 
compared with the  range $\sim 10^{24\mbox{-}28} \mathrm{g}$ probed by the 
pulsar terms.

\begin{figure}
  \includegraphics[width=80mm]{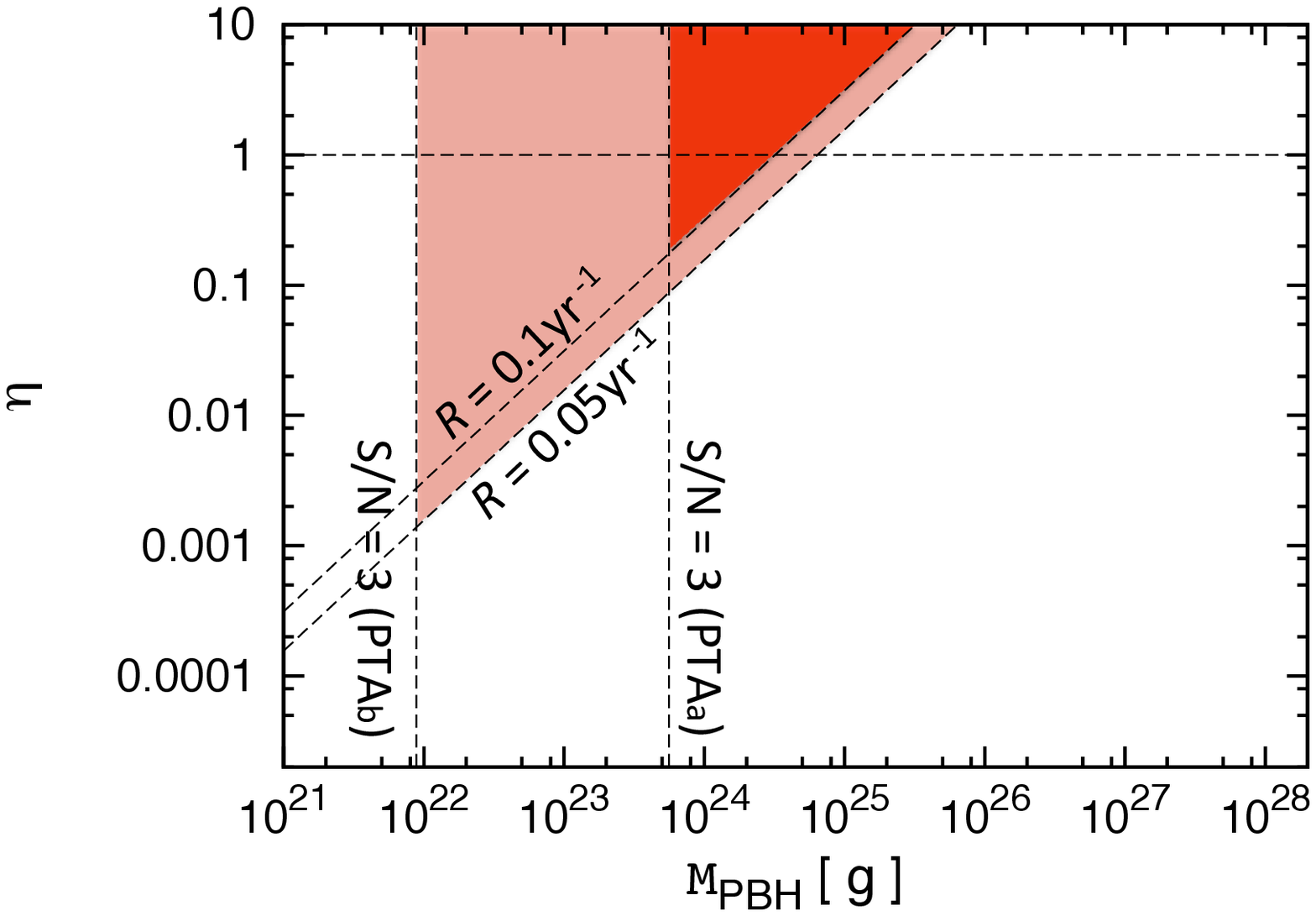} 
  \includegraphics[width=80mm]{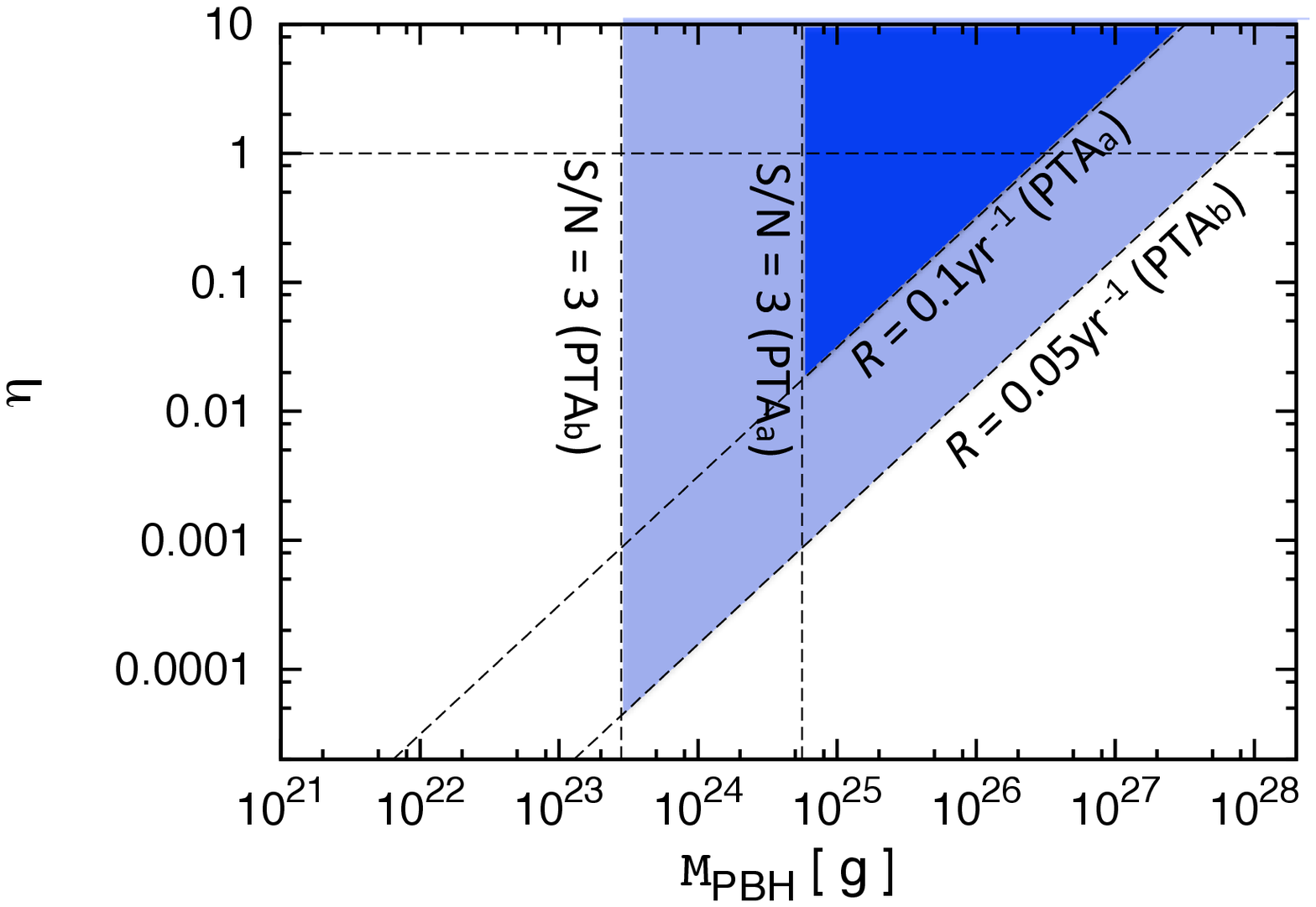} 
  \caption{Possible constraints on the primordial black hole (PBH) abundance 
($\eta \equiv \rho_\mathrm{PBH} / \rho_\mathrm{DM}$) using the two PTAs (the 
dark-colored regions for PTA$a$, and the light ones for PTA$b$). As in Fig.2, the top (bottom) panel shows the case with the Earth 
(pulsar) terms.  We put $V = 350 \mathrm{km/s}$, $|\cos \theta | = 0.58$, and $\rho_\mathrm{DM} = 0.011M_\odot 
\mathrm{pc}^{-3}$. The lines with the slopes  $M_\mathrm{PBH} = \mathrm{const}$, 
and $\eta \propto M_\mathrm{PBH}$  correspond to the conditions (i) and (ii) 
respectively,  with the impact parameter at $b_\mathrm{max}$. }
\end{figure}\label{fig_3}

Fig.3 shows possible constraints on the PBH fraction $\eta$ with the two PTA models.  
Again, the top (bottom) panel shows the case using the Earth (pulsar) terms. The 
detectable region is bounded by two lines.
The condition S/N$>3$ sets the minimum  mass of detectable PBH with  the line  
$M_\mathrm{PBH} = \mathrm{const}$ (see Eq.(\ref{condition_i})), and event rate 
determines the maximum one with  $\eta \propto M_\mathrm{PBH}$ 
(Eq.(\ref{condition_ii})). These boundaries can be understood in the following manner.
 For a given observational period $T_\mathrm{obs}$, 
the impact parameter takes the upper limit $b_\mathrm{max}=VT_\mathrm{obs}$ given in 
Eq.(\ref{condition_iii}). Next, when the PBH mass $M_\mathrm{PBH}$ is fixed, the 
signal-to-noise ratio and the  sensitivity to the ratio $\eta$ both take  their 
optimal values at $b=b_\mathrm{max}$, as shown in Eqs.(\ref{condition_i}) and (\ref{condition_ii}).

\if
Besides, S/N and event rate become larger with a longer impact parameter. (See Eq.(\ref{condition_i}) and (\ref{condition_ii}).)  
Thus, both the maximum and minimum mass of PBHs which can be probed is given 
by the scattering with the maximum impact parameter $b_\mathrm{max}$ (Eq.(\ref{condition_iii})). 
The $M_\mathrm{PBH} = \mathrm{const}$, and $ \eta \propto M_\mathrm{PBH}$ lines in Fig.3 correspond to the condition (i), and (ii) for the scattering with the impact parameter $b_\mathrm{max}$, respectively.
\fi
If no event is detected by the PTA$a$ observation, PBH with $10^{24} 
\mathrm{g}$ to $10^{26} \mathrm{g}$ is excluded as a dominant component of dark 
matter ($\eta \lesssim 0.1$).
Moreover, in the case of the PTA$b$ observation, PBH within $10^{22 \mbox{-}28} \mathrm{g}$ could be excluded ($\eta \lesssim 0.01$).

\section{discussions}
Our order-of-magnitude estimation has shown that we might probe PBHs or constrain 
 their abundance with future PTAs.  
Here let us discuss  issues related to  the PBH search.

\subsection{Data analysis and competing noises}
So far we have simply evaluated the signal-to-noise ratios of fly-by events, 
assuming  white noise spectra for timing noises of MSPs.  But  we should pay much attention to whether PBH signals can be clearly identified in their  TOA data.  

The actual timing analysis typically proceeds as follows (see {\it e.g.} \cite{2006MNRAS.369..655H, 2006MNRAS.372.1549E} for the detail).  
First, the measured TOAs are  converted to the pulse emission times in the 
presumed reference frame of each MSP.
In this step, various effects are estimated, such as the propagation delays,   
the ephemerides, 
and the coordinate transformation from the Earth to the pulsar.
The derived time of emission is fitted by a timing model including a pulse frequency and its time derivative.  
The residual is divided into a white noise and un-modeled systematic noises which could include the signal we are seeking. 
A key in our approach to enlarge the detectable range of PBH mass is to consult 
both the coherent Earth terms and the individual pulsar terms. These two are 
assumed to be separated.  

Here, we should be aware of the risk to mistake PBH signals for other effects while processing the TOA data,
which could also lead to misestimation of the parameters of each pulsar. 
To prevent this, more detailed modeling of the PBH signal is required.

With a PTA, we have a large 
number of independent pulsar terms. The impulse accelerations by fly-by PBHs appear in the 
pulsar terms as sparse and isolated signals with finite durations. Therefore, a long-term 
observational campaign  would be quite helpful to 
select sufficiently stable MSPs and examine  the characteristic time profiles of 
the PBH signals.

As mentioned above,  we have assumed that the ordinary timing noise in each TOA is white, 
$\sigma_\mathrm{WN} \propto T^{0}$ ($T$: the observational time span). 
As for non-recycled pulsars, there have been confirmed secular red noises, $\sigma_\mathrm{RN} \propto T^{2 \pm 0.2}$ \citep{2010ApJ...725.1607S}.
Although most of the observed MSPs only have upper limits on the amplitude of the red noise, 
it might be identified by future PTAs to be obstacle in the PBH search. 

\subsection{Competing signals}
Once a higher sensitivity as we have considered is realized, other possible signals also have to be opened up for discussion. 
They can be effective noises for PBH search and vice versa. 

The most promising target of future PTAs is the GW background from merging supermassive black holes.    
The typical magnitude of the estimated dimensionless strain is  $h \sim 10^{-15.5} \times (f/0.1\mathrm{yr}^{-1})^{-2/3}$ \citep{2003ApJ...583..616J, 2003ApJ...590..691W}.  
This corresponds to the amplitude of the timing residual of  
$s_{f,\mathrm{GW}} \sim 4.1 \mathrm{ns} \times (f/0.1\mathrm{yr}^{-1})^{-5/3}$ (for the translation from $h$ to $s_{f,\mathrm{GW}}$, see {\it e.g.} \cite{2009MNRAS.394.1945H}), 
which can be comparable to that induced by PBHs (see Eq.(\ref{timing_residual})). 
Given the effective sensitivities of the pulsar and the Earth terms, the latter would be more vulnerable to the 
existence of the GW background.  But, using the angular patterns of the Earth terms,  we can, in principle, separate 
the two competing signals, as follows \citep{2007ApJ...659L..33S}.
A PBH signature in the Earth terms  will have a dipole pattern ($l=1$) whose 
direction is determined by the acceleration vector. 
On the other hand, the GW background will have multiple modes starting from the quadrupole ($l=2$). 

One may suspect that floating planets \citep{Sumi:2011} could also become an effective noise  
because of the similar masses to PBHs we are interested in.   
There are expected to be a few times more floating planets than stars. 
The anticipated encounter rate with a pulsar is 
$\sim 10^{-(7\mbox{-}8)} \mathrm{yr}^{-1} \times (b/740\mathrm{AU})^2 $, 
which is much smaller than PBH constituting a dominant fraction of dark matter (see Eq.(\ref{event_rate})). 
Thus, we conclude that floating planets would not be problematic for our approach.    

\subsection{Possible refinements}
Finally let us discuss possible improvements for our estimation of the detectable region of the PBH parameters.  

In Fig.2 and Fig.3, we have fixed $V = 350 \mathrm{km/s}$, $|\cos \theta | = 0.58$, and $\rho_\mathrm{DM} = 0.011 M_\mathrm{\odot} \mathrm{pc}^{-3}$.
These values correspond to the observed rms velocity for halo dark matter relative to the solar system, 
the mean projection factor for isotropic scatterings, 
and the local dark matter density near the solar system, respectively. 
These treatments should be regarded as a zeroth order approximation, and  we 
can refine the 
present analysis in a  more realistic manner.

For example, while the motions of the MSPs would be typically  close to the Galactic rotation, 
 those of PBHs would be more isotropic. 
Then  the mean values for the projection factor
 $\cos \theta$ and the relative velocity $V$ would depend on the Galactic 
position of each target MSP. In  addition, the densities of the PBHs and the 
MSPs would also depend on the Galactic position.  These would affect the 
estimation of the  event rate and the strategy for data analysis.

Future facilities like SKA could determined the distance of MSPs up to $13 \mathrm{kpc}$ with $\lesssim 20 \%$ accuracy \citep{Smits:2011}. 
Multiple detections using the pulsar terms could potentially give us more detailed informations of the PBHs, such as their density and velocity distributions in the whole Galaxy, 
which can not be reached only using the Earth terms. 


\section*{Acknowledgments}
This work is supported in part by the JSPS fellowship for research abroad, the JSPS grants, Nos.20740151 and 24540269. 
 KK thanks P. M$\mathrm{\acute{e}}$sz$\mathrm{\acute{a}}$rose for useful discussions.

\include{ref}

\label{lastpage}

\end{document}

%% file: ref.tex
\bsp